\documentclass[aps,nofootinbib,twocolumn,floatfix]{revtex4} 
\usepackage{graphicx}
\input epsf
\newcommand{\sfig}[2]{
\includegraphics[width=#2]{#1}
        }
\newcommand{\Sfig}[2]{
    \begin{figure}[t] 
    \sfig{#1.eps}{\columnwidth}
    \caption{{\small #2}}
    \label{fig:#1}
    \end{figure}
}

\newcommand{\rf}[1]{\ref{fig:#1}}

\def\vl{{\bf l}}
\def\vt{\mbox{\rm \boldmath$\theta$\unboldmath} } 

\def\Mpc{\,{\rm Mpc}}

\def\cmm2{{\,\rm cm^{-2}}}
\def\cm2{{\,{\rm cm}^2}}
\def\cmm3{{\,{\rm cm}^{-3}}}
\def\gcmm3{{\,{\rm g\,cm^{-3}}}}

\def\VEV#1{\left\langle #1\right\rangle}

\def\fun#1#2{\lower3.6pt\vbox{\baselineskip0pt\lineskip.9pt
  \ialign{$\mathsurround=0pt#1\hfil##\hfil$\crcr#2\crcr\sim\crcr}}}

\def\beq{\begin{equation}}
\def\eeq{\end{equation}}
\def\bea{\begin{eqnarray}}
\def\eea{\end{eqnarray}}

\newcommand{\ec}[1]{Eq.~(\ref{eq:#1})}

\newcommand{\eql}[1]{\label{eq:#1}}

\newcommand{\pderiv}[2]{\frac{\partial#1}{\partial#2}}
\newcommand{\Cl}[2]{C_{#2}(#1)}
\newcommand{\Clobs}[2]{C^{\rm obs}_{#2}(#1)}
\newcommand{\Cmat}[1]{\mathsf{C}_{#1}}
\newcommand{\fsky}{f_{\rm sky}}
\newcommand{\ngal}{\nu^{\rm gal}}
\def\Msol{M_{\odot}}

\def\MAHbias{\overline{b^M}}
\def\Mmin{M_{\rm min}}

\begin{document}

\title{Combining Weak Lensing Tomography with Halo Clustering to Probe Dark Energy}

\author{Charles Shapiro$^{1,2}$, Scott Dodelson$^{3,4,2}$}

\affiliation{$^1$Department of Physics, The
University of Chicago, Chicago, IL~~60637-1433}
\affiliation{$^2$Kavli Institute for Cosmological Physics, The
University of Chicago, Chicago, IL~~60637-1433}
\affiliation{$^3$Center for Particle Astrophysics, Fermi National
Accelerator Laboratory, Batavia, IL~~60510-0500}
\affiliation{$^4$Department of Astronomy \& Astrophysics, The
University of Chicago, Chicago, IL~~60637-1433}

\date{\today}
\begin{abstract}

Two methods of constraining the properties of dark energy are weak lensing tomography and cluster counting.
Uncertainties in mass calibration of clusters can be reduced by using 
the properties of halo clustering (the clustering of clusters). However,
within a single survey, weak lensing and halo clustering probe the same density fluctuations. We explore the question of whether this
information can be used twice -- once in weak lensing and then again in halo clustering to calibrate cluster masses -- or whether the combined dark energy constraints are {\it weaker} than the sum of the individual constraints.
For a survey like the Dark Energy Survey (DES), we find that the cosmic shearing of source galaxies at high redshifts is indeed highly correlated with halo clustering at lower redshifts.  Surprisingly, this correlation does not degrade cosmological constraints for a DES-like survey, and in fact, constraints are marginally improved since the correlations themselves act as additional observables.
This considerably simplifies the analysis for a DES-like survey: when weak lensing and halo clustering are treated as independent experiments, 
the combined dark energy constraints (cluster counts included) are accurate if not slightly conservative.
Our findings mirror those of Takada and Bridle, who investigated correlations between the cosmic shear and cluster counts.

\end{abstract}
\maketitle

\section{Introduction}

Weak lensing tomography~\cite{Hu:1998az,Huterer:2001yu,Hu:2002rm,Abazajian:2002ck,Refregier:2003ct,Ishak:2003zw} 
and cluster counting~\cite{Wang:1998gt,Haiman:2000bw,Holder:2001db,Hu:2002we,Majumdar:2002hd}
 are complimentary methods of constraining dark energy parameters.  Weak lensing tomography (WLT) measures how the images of galaxies at various redshifts are sheared by large scale structure, while cluster counting (CC) simply counts the number of massive clusters as a function of redshift.
Both techniques exploit dark energy's geometric effects and its effects on large scale structure growth.  The spatial clustering of clusters, or halo clustering (HC), provides free additional information that can break degeneracies in cluster surveys~\cite{Majumdar:2002hd,Lima:2004wn}.  Although halo clustering is by far the weakest of the three probes, it becomes more valuable as systematic uncertainties (e.g. cluster mass calibration) degrade dark energy constraints.  Future surveys will combine these techniques.

Since both the cosmic shear and the presence of clusters arise from fluctuations in the cosmic mass density field, we can expect their signals and error bars to be correlated.  Intuitively, if a survey or combination of surveys observes a patch of sky with more clusters than average, we should expect the patch to have an above average cosmic shear signal.
Hence, an analysis that treats lensing and clusters as independent observables will double-count information and may potentially overstate the bounds on cosmological parameters.  On the other hand, 
the cross-correlation between lensing and clusters is an additional observable that can work to improve the parameter fit.  Thus, combining lensing and clusters properly is important: treating them as independent could mean that we have artificially small error bars or that we are 
throwing away valuable information - both disturbing possibilities!

Several authors have investigated WLT/CC correlations, finding that they arise primarily from the small scale power that individual halos contribute to the mass power spectrum.  Approximating the halo number density as a smooth field related linearly to the cosmic mass density, Fang and Haiman~\cite{Fang:2006dt} find negligible covariance between cosmic shear and total cluster counts.  
Takada and Bridle~\cite{Takada:2007fq} further take into account the structure and discrete nature of halos; 
they find a correlation between cluster counts and the shear power spectrum on arcminute scales of about 10\% 
(higher or lower depending on the minimum halo mass used).

We focus here on WLT/HC correlations: these should occur
since shear maps and halos trace the same field of projected matter fluctuations (on scales larger than the average halo spacing).  In other words, a local overdensity of mass tends to create a shear peak and an abundance of halos along the same line of sight.  This correlation turns out to be significantly larger than the WLT/CC correlation, and it warrants a closer look at how halo clustering and weak lensing tomography should be combined.  The WLT/HC correlation is most important in the regime where halo clustering is adding significant constraints to dark energy parameters; we will be working in that regime.

In this paper, we calculate the WLT/HC cross-correlation and determine its effect on a survey's ability 
to measure dark energy parameters.  
We restrict ourselves to large angular scales so that halo clustering is related to the mass overdensity by a linear bias and 
we can ignore the effects of individual halos on the shear power spectra.
We focus on projections for the Dark Energy Survey (DES) and
find that including the full covariance does not degrade the projected constraints; in fact
these tighten at the percent level.

\section{Weak Lensing and Halo Clustering}
Divide a survey region into redshift bins with $z_i$ denoting the minimum redshift of the $i$th bin.  In this paper, we consider 10 bins of equal width up to $z$=1.  Let $\chi(z)$ denote the comoving distance to an object at redshift $z$ so that $\chi_i\equiv\chi(z_i)$ (we will often use $\chi$ and $z$ interchangeably).
The weak lensing convergence is a good computational proxy for shear; in a given redshift bin, the convergence is a function of angular position on the sky and can be expressed as a weighted integral of the density along the line of sight:
\beq
\kappa_i(\vt)=\int d\chi W_{({\rm Lens},i)}(\chi) \delta[\chi\vt,\chi]
\eeq
where the weighting function is
\beq
W_{({\rm Lens},i)}(\chi)=
\frac{3\Omega_m H_0^2}{2\ngal_i}(1+z)\chi\int_{\max(\chi,\chi_i)}^{\chi_{i+1}}d\chi_s \frac{\chi_s-\chi}{\chi_s}\, \frac{d\ngal}{d\chi_s} \\
\eeq
for $\chi\leq\chi_{i+1}$ and 0 otherwise \cite{Takada:2003ef}.  Here $\ngal_i$ is the expected galaxy density in bin $i$ which we
parametrize as
\beq
\frac{d\ngal}{dz}=\ngal_*\frac{z^2}{2z_*^3}\,e^{-z/z_*}
\eeq
The distribution has $z_{\rm mode}=2z_*$ and $z_{\rm median}=2.67z_*$ ; for DES we use $z_*=0.254$ and $\ngal_*=15/{\rm arcmin}^2$.

We approximate the local overdensity of halos in the $i$th redshift bin as another line-of-sight integral,
\beq
h_i(\vt)\equiv\frac{\delta n_i(\vt)}{\bar{n}_i}=
\int d\chi W_{({\rm Halo},i)}(\chi) \delta[\chi\vt,\chi],
\eeq
where $\bar{n}_i$ is the average halo density in bin $i$.  We are counting all halos more massive than a threshold $\Mmin=10^{14.2}\Msol/h$, and we've assumed that the overdensity of halos of a given mass is proportional to $\delta$ (valid on large scales).  The halo weighting function is
\beq
W_{({\rm Halo},i)}=\cases{ \chi^2\,\MAHbias\!(\chi) & $\chi_i<\chi<\chi_{i+1}$\cr
			0 & otherwise}
\eeq
where the {\it mass-averaged halo bias} $\MAHbias\!(\chi)$ depends on both redshift and our $\Mmin$:
\beq
\MAHbias(z)\equiv\frac{\int_{M_{\rm min}}^\infty dM\,b(M,z)\,\frac{dn}{dM}(z)}{\int_{M_{\rm min}}^\infty dM\,\frac{dn}{dM}(z)} .
\eeq
We compute $\MAHbias$ using the halo mass function $dn/dM(z)$ of Jenkins et al. \cite{Jenkins:2000bv} and the halo bias model $b(M,z)$ of Sheth and Tormen \cite{Sheth:1999mn}.

Let $u_I$ be one of our observables with the subscript denoting the data type and redshift bin, 
e.g. $u_I=\kappa_i$ for $I=({\rm Lens},i)$.  Then
\beq
u_I(\vt)=
\int d\chi W_I(\chi) \delta[\chi\vt,\chi]
.\eeq
Using the flat-sky approximation, decompose the observables $u_I$ into 2D Fourier modes. Then we define
\beq
\VEV{u_I(\vl)u_J({\bf l'})} = (2\pi)^2\delta^2(\vl-{\bf l'})\Cl{l}{IJ}
\eeq
where under the Limber approximation, 
\beq
\Cl{l}{IJ} = \int \frac{d\chi}{\chi^2} W_{I}(\chi)W_{J}(\chi)P_{IJ}\left(k=\frac{l}{\chi};\chi\right)
\eql{CIJ}
\eeq
The matter power spectrum $P_{IJ}(k)$ depends on the data types.  
We use a linear spectrum to compute halo-halo and halo-lens terms and a nonlinear spectrum to compute lens-lens terms\footnote{In the formalism we've chosen, the proper treatment of the halo-lens term is ambiguous, but our results are insensitive to our matter power spectrum choice.}.  Our matter power spectra are computed using the fitting formula of Eisenstein and Hu \cite{Eisenstein:1997jh} and the nonlinear scaling of Smith et al. \cite{Smith:2002dz}.  Note that the weighting functions $W_I$ imply that
\bea
\VEV{h_i h_j}&=&0\mbox{ for } i\neq j \\
\VEV{\kappa_i h_j}&=&0\mbox{ for } i<j
.\eea
These follow from the Limber approximation and should be valid on scales $l>10$.

\Sfig{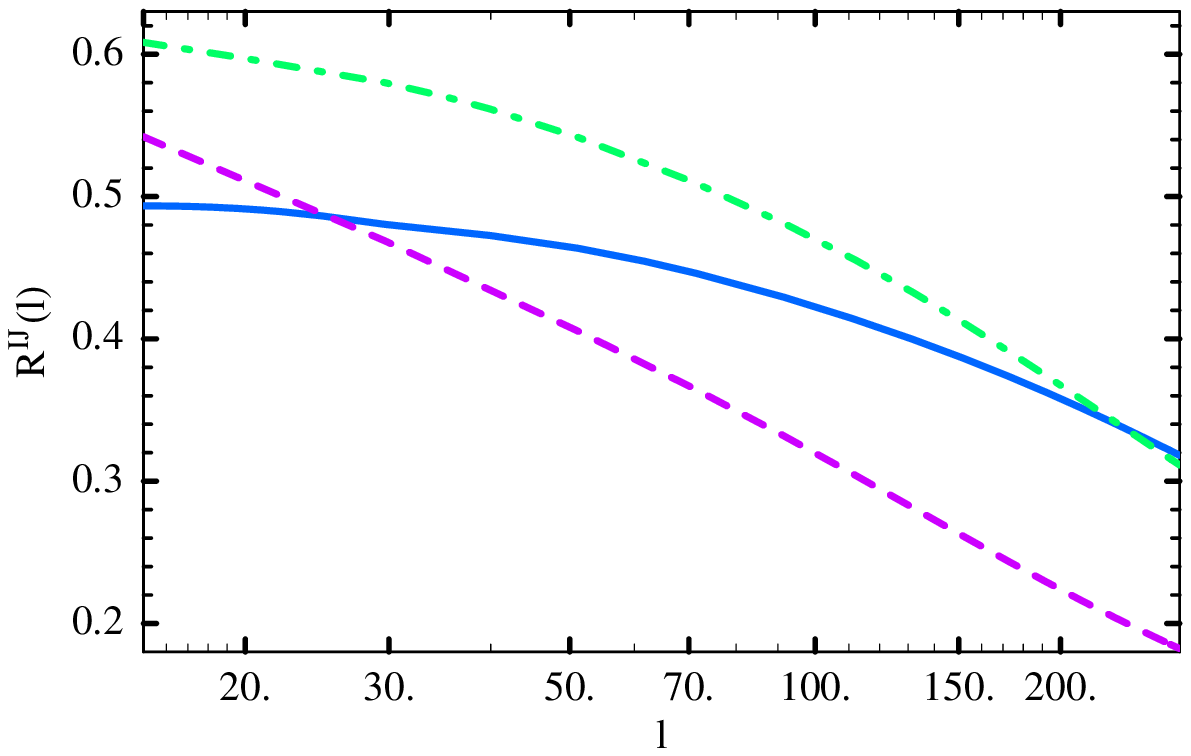}{Sample correlation coefficients; for a given lensing bin, 
the most strongly correlated halo bin is shown.  All bins have width $\Delta z=0.1$, and the average redshifts of the bins (Lens, Halo) are as follows: solid blue (0.95, 0.25); dot-dashed green (0.65, 0.15); dashed purple (0.35, 0.05).}

To assess the degree to which lensing and halo clustering are correlated, it is useful to consider
the correlation coefficient
\beq
R_{IJ}\equiv \Cl{l}{IJ}/\sqrt{\Cl{l}{II}\Cl{l}{JJ}}
.\eeq
If lensing were completely uncorrelated with halo clustering, the correlation 
coefficients $R_{({\rm Lens},i)({\rm Halo},j)}$ would vanish; if they
measured identical quantities, the coefficient would be unity.
Figure~\rf{covar_coeff_signal} shows that on large scales, the lensing of high redshift background galaxies is highly correlated with clustering at intermediate 
redshifts. It is therefore possible that tighter constraints anticipated from including halo clustering are
illusory: perhaps this information has already been used once -- in weak lensing tomography -- and can't be double-counted.  In that case, it would not be appropriate
to blindly combine the constraints from (CC+HC) with those from WLT.

\section{Impact on Dark Energy Projections}

To assess the impact of the correlations depicted in Fig.~\rf{covar_coeff_signal}, we compute
the Fisher matrix for weak lensing tomography and cluster counting supplmented by halo clustering. We choose a standard set of cosmological parameters in a flat universe:
\newcommand{\td}[1]{\hspace{.1cm}$#1$\hspace{.1cm}}
\begin{center}
\begin{tabular}{c|ccccccc}
Parameter & \td{\Omega_mh^2}&\td{\Omega_bh^2}&\td{\Omega_{\rm de}}&\td{w_0}&\td{w_a}&\td{n}&\td{\delta_\zeta^2}\\
\hline
Fiducial & .14&.024&.73&-1&0&1&$2.57\times10^{-9}$\\
Prior & 1\%&1\%&-&-&-&1\%&1\%\\
\end{tabular}
\end{center}
where $h$ is the Hubble constant in units of
$100$ km/sec/Mpc; $\Omega_{\rm de}$ is the ratio of the dark energy density to the critical density (and
similarly for the total matter and baryon densities); $n$ is the spectral index of primordial scalar fluctuations, and $\delta_\zeta$ is their amplitude\footnote{In practice, we vary $\ln \delta_\zeta^2$.  The prior is defined at $k=0.05\,\Mpc^{-1}$.} at $k=0.002\,\Mpc^{-1}$; the dark energy equation of state is taken to be
$w=w_0+w_a z/(1+z)$.  We impose Gaussian priors expected from future CMB surveys. Additionally, we include (with no priors) the cluster mass calibration parameters $A$ and $n_A$ defined by Lima and Hu \cite{Lima:2004wn}.

The Fisher matrix for our parameter set is, in the standard approximation, a sum of matrices for our three observables:
\beq
F^{\rm Total} = F^{\rm CC} + F^{\rm HC} + F^{\rm WLT}
.\eql{fisher_approx}
\eeq
The full cluster Fisher matrix breaks nicely into a CC and HC term, as in the above equation \cite{Lima:2004wn}.  However, we know from the results of \S II (e.g. Fig.~\rf{covar_coeff_signal}) and previous work (\cite{Fang:2006dt,Takada:2007fq}) that lensing and clusters are not completely independent observables, and therefore, that the above approximation is flawed.  To account for the correlation between halo clustering and cosmic shear, we generalize $F^{\rm Total}$ to
\beq
F^{\rm Total} = F^{\rm CC} + F^{\rm HC+WLT} + \tilde F^{\rm WLT}
\eql{fisher_full}
\eeq 
where $\tilde F^{\rm WLT}$ contains lensing information on scales smaller than can be probed with halo clustering.

The observed spectra $\Clobs{l}{}$ are the sum of the {\it 
signals} given by \ec{CIJ} and Poisson noise (approximated as Gaussian):
\beq
\Clobs{l}{IJ}\equiv\Cl{l}{IJ} + \delta_{IJ}N_I
\eeq
where $\delta_{IJ}$ is a Kronecker delta and
\beq
N_I = \left\{
\begin{array}{ll}
\gamma_*^2/\ngal_i & \mbox{\rm for } I=({\rm Lens},i) \\
1/\bar{n}_i & \mbox{\rm for } I=({\rm Halo},i) 
\end{array}\right.
\eeq
Here, $\gamma_*$ is the RMS intrinsic shear of the source galaxies; for a DES-like survey, $\gamma_*\approx0.25$ \cite{Hoekstra:1999pu}.  
Let $\Cmat{l}$ be a matrix with elements $\Clobs{l}{IJ}$.  Then for a set of parameters $p_\alpha$,
\beq
F^{\rm HC+WLT}_{\alpha\beta}=
\fsky \sum_{l} \frac{2l+1}{2}\;
{\rm Tr}\left[\pderiv{\Cmat{l}}{p_\alpha}\Cmat{l}^{-1}\pderiv{\Cmat{l}}{p_\beta}\Cmat{l}^{-1}\right]
\eql{fisher_general}
\eeq
where $\fsky$ is the fraction of sky covered by the survey, and the trace sums over all redshift bins and data types\footnote{We check that our numerical derivatives converge with decreasing step size.  Ultimately, we use two-sided derivatives with steps that are 4\% of the parameters or 0.04, 0.01 and 0.01 for $w_a$, $A$ and $n_A$ respectively.}.
Similarly, we can write $\tilde F^{\rm WLT}$ this way, summing only over the lensing part of  $\Cmat{l}$.  Note that $F^{\rm HC+WLT}$ is a sum over $l$ from $l_{\min}$ to $l_{\rm mid}$, while $\tilde F^{\rm WLT}$ is summed from $l_{\rm mid}$+1 to $l_{\max}$.  We choose $l_{\min}$=20 and $l_{\rm mid}$=200 so that the flat-sky approximation and our linear halo bias assumption are valid.  As prescribed by Rudd et al. \cite{Rudd:2007zx}, we modestly set $l_{\max}$=1000 since baryonic effects on smaller scales are not completely understood.

If halo clustering were a completely independent observable from cosmic shear, then the matrices $\Cmat{l}$ would be block diagonal with no terms in the (HC,WLT) region.  In that case, \ec{fisher_full} reduces to  \ec{fisher_approx}.  We compute $F^{\rm Total}$ for both cases, and since we are not investigating CC correlations, our cluster counting term is always given by
\beq
F^{\rm CC}_{\alpha\beta} = \sum_{i} \pderiv{m_i}{p_\alpha}\pderiv{m_i}{p_\beta}\frac{1}{m_i}
\eeq
where $m_i$ is the total number of clusters in the $i$th redshift bin with $M\ge M_{\rm min}$.  Note we've assumed that Poisson noise is the dominant source of scatter for cluster counting.

After computing the Fisher matrix, we use it to forecast 1$\sigma$ errors for the dark energy parameters after marginalizing over all other parameters.  We also compute the dark energy figure of merit (FoM), defined as the inverse area of the 95.4\% error ellipse for ($w_0,w_a$) \cite{Albrecht:2006um}.  It turns out that the HC/WLT correlations have very little impact on the final parameter determination and even bring a slight improvement for DES.  Numbers quantifying this small impact are given in Fig.~\rf{dFoM_contours} and Table \ref{tab:results}.  We remind the reader that Fig.~\rf{dFoM_contours} varies two of several possible survey parameters: $M_{\rm min}$, $z_*$ and $\gamma_*$ are held fixed.
\Sfig{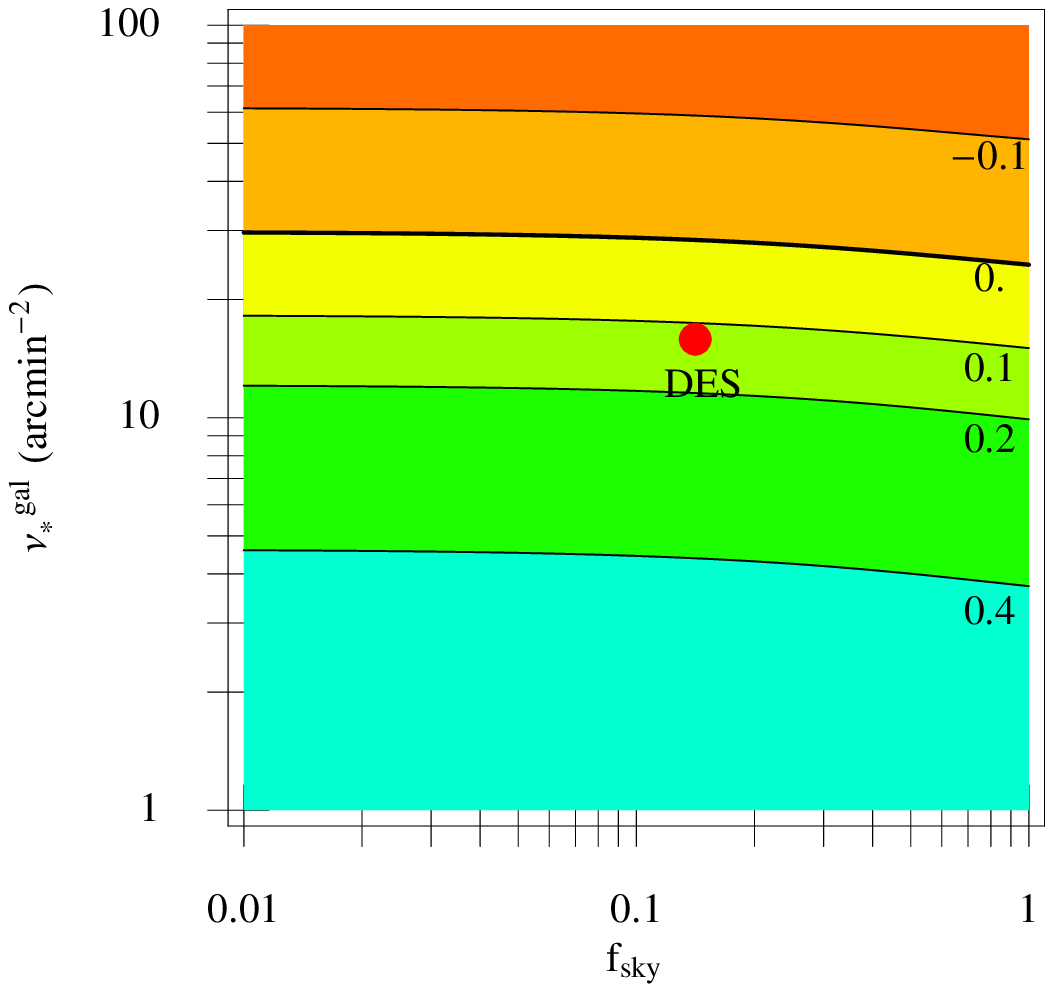}{Fractional change in the dark energy figure of merit once  WLT+HC covariance is included, shown as a function of sky coverage and galaxy density.  Positive values imply that neglecting the covariance results in an underestimation of the FoM (conservative parameter constraints).}

\begin{table}[t]
\begin{tabular}{c|c|c|c|c}
Parameter&No HC& WLT+HC & WLT+HC & Difference \\
&& Correlated & Independent & \\
\hline
$\sigma(w_0)$&0.601&0.461&0.470&-2\%\\
$\sigma(w_a)$&2.15&1.52&1.56&-2\%\\
$\sigma(\Omega_{\rm de})$&0.0387&0.0362&0.0377&-4\%\\ \hline
FoM&0.485&1.15&1.01&+14\%\\
\end{tabular}
\caption{1$\sigma$ error forecasts and dark energy figure of merit for a DES-like survey ($\fsky$=0.12, $\ngal_*$=15/arcmin$^{-2}$).  Constraints in the 2nd column combine CC+WLT only.  Columns 3 and 4 show constraints from CC+WLT+HC, treating WLT+HC as correlated or as independent observables.  The last column compares columns 3 and 4.}\label{tab:results}
\end{table}

Our results show that cosmological constraints 
can be marginally improved by including the full HC/WLT covariance since the HC/WLT correlation itself acts as an additional observable.  Whatever degradation occurs from acknowledging that lensing and clusters are correlated may be offset by our ability to predict and measure the correlation.  We note that correlations do not necessarily imply error degradations in the first place: this would be the case if our set of observables were correlated in such a way that does not mimic their dependence on dark energy parameters.  Mirroring the results of Takada and Bridle~\cite{Takada:2007fq}, we conclude that a DES-like survey that treats halo clustering and 
weak lensing tomography as independent experiments will have accurate if not slightly conservative parameter constraints.  


\section{Conclusions}

The distribution of halos along a given line of sight probes the mass density along that line of sight in ways
very similar to weak lensing. We have quantified this similarlity with the cross-correlation coefficients shown
in Fig.~\rf{covar_coeff_signal}. However, for a survey like DES, projected constraints on dark energy parameters are not degraded by this
cross-correlation (using the same information twice), and it is acceptable to simply add the constraints from weak lensing to those from clusters.  Doing so actually results in slightly conservative parameter constraints.

\begin{acknowledgments}
We thank Simon DeDeo, Sarah Hansen, Wayne Hu, Dragan Huterer, Doug Rudd, and Andrew Zentner for useful discussions.  We thank our anonymous referee for constructive criticism our manuscript.  Thanks also to Antony Lewis, Zhaoming Ma and Eduardo Rozo for assistance with code debugging.  Our matter power spectra were calculated using code donated by Andrew Zentner.  CS was supported in part by the Department of Energy and by the Kavli Institute for Cosmological Physics at the University of Chicago through NSF grant PHY-0114422, NSF grant PHY-0551142 and an endowment from the Kavli Foundation.  SD is supported by the US Department of Energy.
\end{acknowledgments} 

\bibliography{lens_cluster}
\end{document}